\documentclass[aps,prx,superscriptaddress,floatfix,twocolumn]{revtex4}

\usepackage{graphicx}
\usepackage{epsfig}
\usepackage[usenames]{xcolor}

\begin{document}

\title{Electronic Pair-Binding and Hund's Rule Violations in Doped $C_{60}$}

\author{Hong-Chen Jiang}
\affiliation{Stanford Institute for Materials and Energy Sciences, SLAC National Accelerator Laboratory and Stanford University, Menlo Park, CA 94025, USA}

\author{Steven Kivelson}
\affiliation{Department of Physics, Stanford University, Stanford, CA 94305, USA}

\date{\today}

\begin{abstract}%
We calculate the electronic properties of the $t$-$J$ model on a $C_{60}$ molecule using the density-matrix renormalization group and show that Hund's first rule is violated and that for an average of three added electron per molecule, an effective attraction (pair-binding) arises for intermediate values of $t/J$. Specifically,  it is energetically favorable to put four electrons on one $C_{60}$ and two on  a second rather than putting three on each. Our results show that a dominantly electronic mechanism of superconductivity is possible in doped $C_{60}$.
\end{abstract}


\maketitle

Interest in the superconductivity of the alkali-metal doped $C_{60}$ compounds (fullerides) \cite{Hebard1991, Rosseinsky1991, Holczer1991} derives in part from their status as a new class of superconductors with large values of the superconducting critical temperatures $T_c$.\cite{Gunnarsson1997} There has been a great effort over the last two decades to characterize and understand both the normal-state and the superconducting properties of fullerides. A source of renewed interest in these systems is the surprising indication of magnetism derived from strong electron-electron repulsions 
  in crystals in which the $C_{60}$-$C_{60}$ distance is modestly expanded \cite{Prassides2008,Takabayashi2009,Prassides2010,Alloul2010,Alloul2014} Several examples are now known where this kind of expansion first leads to superconductivity with a dome-shaped $T_c$, followed by a Mott insulating state. A superconducting dome proximate to an antiferromagnetic Mott insulating state is a hallmark of strong electron correlations in the high temperature superconducting cuprates and organic charge-transfer salts; its appearance in alkali-   doped $C_{60}$ suggests that electron correlations are crucial in understanding the superconductivity in these materials, as well.

Although most theoretical work has focused on  phonon mechanisms, a dominantly electronic mechanism has 
also been considered. In particular, it was argued in Refs. \cite{Kivelson1991, Kivelson1991p2} that the special geometry of the $C_{60}$ molecule (that of a truncated icosahedron - or more colloquially a soccer ball) permits subtle intra-molecular electronic correlation effects that give rise   to an effective attraction (i.e., positive pair-binding energy) between doped electrons and violations of Hund's first rule. This conjecture was supported by extrapolating second-order perturbative calculations for the one-band Hubbard model on the $C_{60}$ structure  to intermediate values of $U/t$ (where, strictly speaking, low order perturbation theory is not justified). These inferences were also supported\cite{Kivelson1992} by exact diagonalization (ED) studies of smaller ``Hubbard molecules'' -- especially the somewhat analogous 12 site truncated tetrahedron. However, various  later numerical studies\cite{Murthy1992, Goff1992, Baskaran1992, Scalettar1993, Goff1993, Krivnov1994, Zhang1995, Sondhi1995, Campbell1995} of the $C_{60}$ problem gave inconclusive and conflicting results.  Most significantly, the best existing quantum Monte Carlo (QMC) calculations \cite{Kallin2005} on the same system suggested significant failures of the extrapolated perturbation theory;  in particular, the QMC results seemingly support the validity of Hund's rule and an absence of pair binding.

In order to resolve the issues of principle, we use density-matrix renormalization group (DMRG)\cite{White1992DMRG, DMRG_RMP, DMRG_2D} to investigate the ground state properties of the $t$-$J$ model on a single $C_{60}$ molecule, including the magnetic properties and electronic pair-binding energy of doped electrons. The $t$-$J$ model [see Eq.(\ref{Eq:TJModel})] is a simplified model of doped $C_{60}$ which, in common with the Hubbard model, can plausibly be assumed to capture  the most significant correlation effects of the system. Moreover, since the $t$-$J$ model is defined on a significantly smaller Hilbert space, it is much less numerically demanding than the Hubbard model.
 Our most important conclusion is that electronic pair-binding (an effective attractive interaction) arising from a purely
  electron-electron repulsions is a now established feature of the $t$-$J$ model for a finite interval of the dimensionless parameter, $t/J$.  In particular, it is energetically favorable to add four electrons to one $C_{60}$ molecule and two to a second rather than to add three electrons to each of  two $C_{60}$ molecules --  i.e. there is a positive pair binding energy.  In addition, we find that Hund's first rule is violated; the ground state is the state of minimal total spin rather than maximal.  For instance, we find that the ground states with two and four doped electrons has spin zero while the groundstate with three doped electrons has spin 1/2.

In the noninteracting limit, the electronic structure of the $C_{60}$ molecule is well known\cite{Haddon1986}, and the electronic states can be labeled according to the irreducible representations of the icosahedral group. 
The neutral $C_{60}$ molecule has a unique ground-state and a substantial gap between the filled and empty orbitals.  The lowest unoccupied molecular orbitals are the threefold-degenerate $t_{1u}$ orbitals, whose degeneracy is an important property of the molecule. For many purposes, the $C_{60}$ molecule can be approximated as a sphere, and the  $t_{1u}$ orbitals can then be thought of as p-orbitals with   ``angular momentum'' $L=1$. The electrons donated by the alkali-metal atoms to the $C_{60}$ molecule enter the threefold-degenerate $t_{1u}$ orbitals. In the presence of orbital degeneracy and weak interactions, the Hund's rules can be derived perturbatively,\cite{Kivelson1991} where Hund's first rule is that the exchange energy is minimized when the molecular state has the highest possible total spin and the second rule is that it has the highest total orbital angular momentum compatible with the first rule. These rules,
 if they applied, would imply that the ground state has a total spin $1$ and ``angular momentum'' 1 when doped with two or four electrons, and a total spin $3/2$ and ``angular momentum'' 0 when doped three electrons.

\section{Verifying convergence}

The steps we have taken to test that all our DMRG results for the $t$-$J$ model on the $C_{60}$ molecule have converged with high accuracy are described in detail in Appendix \ref{Sec:C60TJModel}.  To get convincing results we have had to keep extraordinarily large numbers of DMRG states, and to iterate the DMRG a very large number of times.  However, by doing this we have been able to obtain  results that we are confident which have converged to the exact answer. The Hubbard model on a single $C_{60}$ would, presumably, require keeping an even larger number of states; we have not currently succeeded in obtaining clearly converged results for this more difficult problem.

As a further test of the reliability of our simulation, we have benchmarked the DMRG method on the one-band Hubbard model on the $C_{20}$ molecule. (See Appendix \ref{Sec:C20Hubbard} for details.) We find that the DMRG results converge very rapidly to the exact diagonalization (ED) results, even with a relatively small number of DMRG states, and in particular gets values for both the ground state energy and the pair binding energy that are more accurate than those obtained using QMC. (See Fig.\ref{Fig:C20Hubbard} in Appendix \ref{Sec:C20Hubbard} for details.)

\section{$t$-$J$ model on 
a  $C_{60}$ molecule}%
We now investigate the ground state properties of the $t$-$J$ model on the single $C_{60}$ molecule using DMRG. 
The $t$-$J$ model Hamiltonian on the $C_{60}$ molecule  is 
\begin{eqnarray}
H = \sum_{\langle ij\rangle \sigma}t_{ij}\left( c^+_{i\sigma}c_{j\sigma} + h.c.\right) + J\sum_{\langle ij\rangle} \left( \vec{S}_i\cdot \vec{S}_j -\frac{1}{4}n_in_j\right), \label{Eq:TJModel}
\end{eqnarray}
where $\langle ij\rangle$ are nearest-neighbor (NN) sites, $c^+_{i\sigma}$ ($c_{i\sigma}$) is a fermionic creation (annihilation) operator with spin-$\sigma$ on site $i$, $\vec{S}_i$ is the spin and $n_i=\sum_\sigma c_{i\sigma}^+ c_{i\sigma}$ is the number of holes on site $i$.  The Hilbert space is constrained by the no-double occupancy condition,
$n_i\leq 1$. The sign of the hopping term in  Eq. \ref{Eq:TJModel} is the opposite of the usual convention.  We are interested in ``electron-doped'' $C_{60}$ in which the total number of electrons is $N_e=60+n_e$, where $n_e=0 \ - \ 6$ is the number of ``doped'' electrons added to the neutral $C_{60}$ molecule. However, in deriving Eq. \ref{Eq:TJModel}, we have made a particle-hole transformation, which results in this sign change, and correspondingly it is to be understood that $\sum_i n_i =60-n_e$.  For simplicity, we take all the nearest-neighbor hopping matrix elements 
to be equal to each other, $t_{ij}=t>0$, although in fact their are two inequivalent sets of nearest-neighbor bonds -- those bounding pentagonal plaquettes and those connecting the pentagons. We will define the unit of energy such that $J=1$, and consider the range of parameters $t/J=1 \ - \ 5$, which approximately corresponds to the Hubbard model at $U/t\sim 4t/J=4 \ - \ 20$. For the present DMRG simulation, we employ the standard approach \cite{DMRG_RMP, DMRG_2D} to choose a suitable one-dimensional path over all sites of the $C_{60}$ molecule. We perform up to $100$ sweeps and keep up to $m=12000$ DMRG states with a typical truncation error $\sim 10^{-4}$. This led to excellent convergence for the results that we report here. Extrapolating to $m=\infty$ gives typical fractional errors in the total energy of about $\sim 10^{-3}$.

\begin{figure}
\centerline{
    \includegraphics[height=2.0in,width=3.6in] {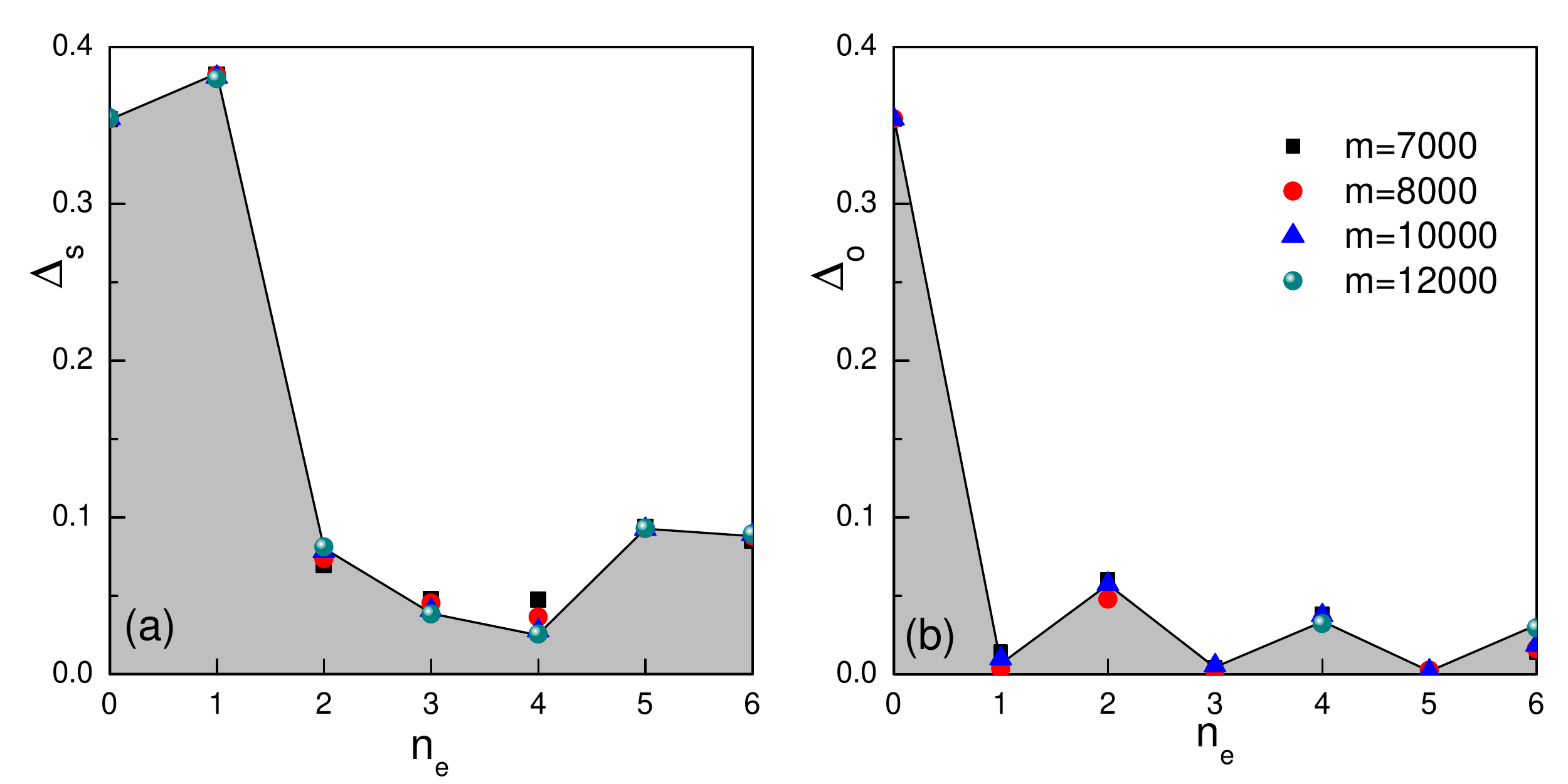}
    }
\caption{(Color online) (a) The ``spin gap'' $\Delta_s$ and (b) orbital gap $\Delta_o$ (defined in Sec. \ref{Hundsrule}) as a function of number of added electrons $n_e$, for  different number of DMRG states $m$. Here $t/J=2$.} \label{Fig:SGap}
\end{figure}

\section{Hund's rule violation}%
\label{Hundsrule}
One of our main observations is that  Hund's first rule is violated for this  range of parameters. 
 The state with the minimal possible total spin has $s_{min}=1/2$ for $n_e$ odd  and $s_{min}=0$ for $n_e$ even, i.e. $\vec S\cdot\vec S \geq s_{min}(s_{min}+1)$ where $\vec{S}=\sum_i \vec{S}_i$ is the total spin operator.  
 The ground-state is generically an eigenstate of total spin, so in testing for violations of Hund's first rule, we measure the ``excess spin,'' $\delta S^2\equiv \langle \vec S\cdot\vec S\rangle -s_{min}(s_{min}+1)$, which is zero in the minimal spin state, and satisfies the inequality $\delta S^2\geq 2$ otherwise. We always find that, for large enough number of kept states, $\delta S^2=0$ to high accuracy ($\delta S^2 < 1.2\times 10^{-2}$); representative data for $t/J=2$  for all values of $n_e$ in the range $0\ - 6$ are shown in Fig. \ref{Fig:STot2SGap}(a). (See Appendix \ref{Sec:C60TJModel} for details.) In addition, we  define the ``spin-gap,'' $\Delta_{s}\equiv E_0(s_{min}+1)-E_0( s_{min})$, where $E_0(S_z)$ is the ground state energy for given value of $z$-component of total spin $S_z$. 
 In any state with more than the minimal spin, $\Delta_{s}=0$, while, baring an accidental degeneracy, in a minimal spin state, $\Delta_{s}>0$. The value of $\Delta_{s}$ as a function of $n_e$ is shown in Fig. \ref{Fig:SGap}(a) for $t/J=2$;  the different colored points represent the results with different numbers of kept states, $m$. (A more complete presentation of the convergence to the $m\to\infty$ limit is shown in Fig.\ref{Fig:STot2SGap} (a).) It is clear that $\Delta_{s}$ is non-zero in all cases, which is independent confirmation of the conclusion that the ground state has the minimum possible spin. For  $n_e=2$, 3, and 4  this represents a violation of Hund's first rule.

In Fig.  \ref{Fig:SGap}(b) we show the ``orbital gap,'' $\Delta_o\equiv E_1(s_{min})-E_0(s_{min})$, where $E_1(S_z)$ is the energy of the first excited for given $S_z$. If the ground state is an orbital singlet, $\Delta_o>0$, while for any orbital multiplet (higher angular momentum) $\Delta_o = 0$. For $n_e$ even, both $\Delta_s$ and $\Delta_0$ are non-zero, implying that the ground states are both orbital and spin singlets.  For odd $n_e$, that $\Delta_s>0$ and $\Delta_o=0$  implies that the ground states have spin 1/2 and are orbital multiplets.   

All these findings are consistent with an analysis\cite{Kivelson1991} in which the ground-states are adiabatically connected to appropriate (symmetry determined) combinations of the non-interacting ground-states:  The states with one or five electrons in a p-orbital  have total spin $s=1/2$ and angular momentum $l=1$ (i.e. an orbitally degenerate minimal spin state consistent with what we find). The states with  two or four electrons can have $s=1$ and $l=1$ (the state favored by Hund's first rule), $s=0$ and $l=2$ (the state favored by Hund's second rule, if the first were to be ignored), or $s=0$ and $l=0$ (i.e. an orbitally non-degnerate minimal spin state consistent with what we find).  The states with  three electrons can have $s=3/2$ and $l=0$ (the state favored by Hund's first rule), or $s=1/2$ and $l=2$ or $l=1$ (either of which is an orbitally degenerate minimal spin state consistent with what we find).  From the weak coupling perspective, the fact that the ground-state is an orbital and spin singlet when $n_e=0$ or 6 appears obvious (corresponding to an empty or full $t_{1u}$ orbital).  However, the fact that the ground states of the $t$-$J$ model have the same symmetries as the non-interacting ground-states even in these cases is a non-trivial observation.  In particular, for $n_e=0$, this is a statement concerning the ground-state of the spin-1/2 Heisenberg model on the $C_{60}$ lattice, a problem which has many interesting features in its own right.\cite{Trugman1992}

\begin{figure}
\centerline{
    \includegraphics[height=2.6in,width=3.4in] {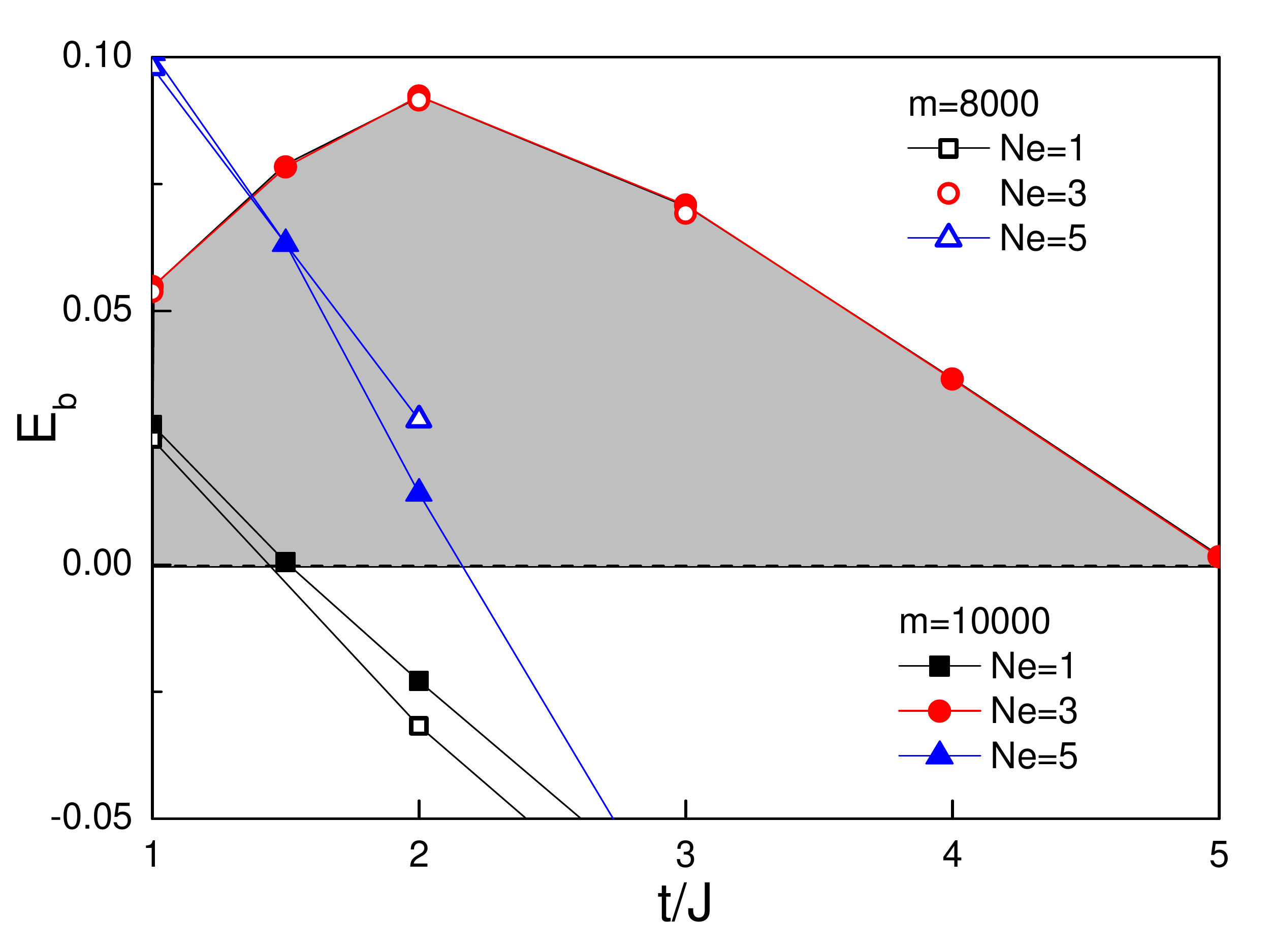}
    }
\caption{(Color online) Electronic pair-binding energy $E_b(n_e)$ as a function of $t/J$ with different $n_e$ for both $m=8000$ and $m=10000$ number of DMRG states. The shaded region and lines connecting the data points are guides to the eye only.} \label{Fig:Eb}
\end{figure}

\section{Pair-binding energy}%
The electronic pair-binding energy is defined as $E_b(n_e)=2E_0(n_e)-E_0(n_e+1)-E_0(n_e-1)$, where $E_0(n_e)$ is the ground-state energy of the system with $n_e$ doped electrons. If we consider a system with an average of $n_e$ doped electrons per molecule, a positive pair-binding energy can be interpreted as an effective attraction between electrons in the sense that it is then energetically favorable to add $n_e+1$ electrons to half the molecules and $n_e-1$ to the other half, rather than to place $n_e$  electrons on every molecule. For $n_e$ even, we always find that the pair-binding energy is negative.  However, for $n_e=1$, 3, and 5, $E_b(n_e)$ is positive for a range of intermediate $t/J$.  This is illustrated in Fig.\ref{Fig:Eb}, which shows $E_b(n_e)$ as a function of $t/J$ for $n_e=1$, $3$ and $5$.  
Importantly, for the whole $t/J$ parameter region we have explored, $E_b(n_e=3)$ is positive, although at our largest value of  $t/J=5$ it is close to zero and appears to be headed to negative values at still larger $t/J$. For $n_e=1$ and $n_e=5$, $E_b$ is positive for small enough $t/J$, but crosses zero and is distinctly negative (corresponding to an effective repulsion between electrons) beyond a critical value of $t/J$. 

For  $t/J\lesssim 1$, it can be plausibly argued that the $t$-$J$ model is unphysical, and in particular has, in effect, microscopically attractive interactions, so the results for $n_e=1$ and 5 are of uncertain physical significance, as pair binding is only seen for $t/J \lesssim 1.5$ and $t/J \lesssim 2$, respectively.  But the pair-binding for $n_e=3$ is manifestly robust in the regime $2 < t/J < 5$, where these concerns do not arise.  (Note that the absence of pair-binding as $t/J \to \infty$ is expected on general grounds.  For $n_e=1$, a rigorous proof exists\cite{chayes} that $E_b \leq 0$ in this limit, as a correlary of a generalized version of Nagaoka's theorem. Under the assumption that the fully spin polarized (Nagoaka) state is the ground-state for large enough $t/J$ for $n_e=3$  it follows that $E_b(3)=0$ at large $t/J$.)

\section{Conclusions}%
In this paper, we have studied the $t$-$J$ model on the $C_{60}$ molecule through DMRG simulation. Several different quantities are calculated, including the ground state energy, spin excitation gaps and electronic pair-binding energy, for $n_e=0 \ - \ 6$ and $1 \leq t/J \leq 5$. In all cases, the ground-state has the minimum possible spin, which for $n_e=2$, 3, and 4 constitutes a violation of Hund's first rule.  Correspondingly, for all $n_e$, there is a non-zero spin gap. The ground-state is an orbital singlet for the even values of $n_e$ and orbitally degenerate for the odd values. For $n_e=3$ we find a positive pair-binding energy for the entire range of $t/J$;  thus, we establish that it is possible that an effective attraction of the sort necessary to mediate superconducting pairing, can arise from purely repulsive electron-electron interactions on a single $C_{60}$ molecule. This establishes an important point of principle.  

Naturally, our results do not address the issue of what differences arise in considering  more realistic (non-zero range) microscopic electron interactions, some of which can be expected to enhance\cite{Sondhi1995, Campbell1995} and others to suppress\cite{Kivelson1992, Murthy1992, Goff1993, Krivnov1994, Zhang1995} pair-binding.  It also leaves open the relevance of our findings to the physical problem of superconductivity in alkalai doped $C_{60}$ where both inter-molecular interactions, and electron-phonon interactions\cite{Gunnarsson1997} must be included in a complete analysis of the problem.   In this context, it is important to note that the purely electronic model we have solved results in precisely the same inversion of Hund's rule that elsewhere\cite{auerbach,tosatti} has been attributed to the effect of Jahn-Teller phonons.  The putative  signatures  of a dynamical Jahn-Teller effect -- including the remarkable recent  experimental observations reported  in Ref.\cite{prassides} -- in most cases depend more on the emergent symmetries of the molecular ground-states,  than on the details of the mechanism that produces these states.  As  there is no distinction in symmetry between the molecular ground-states favored  by the dynamical Jahn-Teller effect and those of the $t-J$ model, unraveling the relative importance of the various contributions to the physics of real materials is likely to be more subtle than was previously believed.

\section{Acknowledgments}%
We thank J. Berlinsky, K. Prassides, S. Chakravarty, H. Yao, P. Z. Tang, M. Mulligan and especially T. P. Devereaux for interesting discussions. HCJ and SAK were supported by the Department of Energy, Office of Science, Basic Energy Sciences, Materials Sciences and Engineering Division, under Contract DE-AC02-76SF00515. The computational
work was partially performed at Sherlock cluster in Stanford.

\appendix
\renewcommand{\thefigure}{S\arabic{figure}}
\setcounter{figure}{0}
\renewcommand{\theequation}{S\arabic{equation}}
\setcounter{equation}{0}

\section{Hubbard model on the $C_{20}$ molecule}\label{Sec:C20Hubbard}

The one-band Hubbard model on the $C_{20}$ molecule is given by the Hamiltonian%
\begin{eqnarray}
H=-t\sum_{\langle ij\rangle \sigma}\left(c_{i\sigma}^+ c_{j\sigma} + H.c.\right) + U\sum_i n_{i\uparrow}n_{i\downarrow}. \label{Eq:Hubbard}
\end{eqnarray}
Here $c_{i\sigma}^+$ is the electron creation operator with spin-$\sigma$ on site $i$, and $n_{i\sigma}=c_{i\sigma}^+c_{i\sigma}$ is the number of electrons with spin-$\sigma$ on site $i$. $t$ is the nearest-neighbor hopping constant and $U$ is the on-site Coulomb interaction. Previous studies \cite{Berlinsky2007, Berlinsky2007p2} using QMC and exact diagonalization methods have found a negative pair-binding energy (repulsive interaction) on this molecule. Moreover, they found that the Hund's rule is obeyed for the corresponding range of parameters $U/t\leq 3$ where the ground state has the maximum values of total spin ranging from spin-$1$ for $20$ electrons through spin-$2$ for $22$ electrons, while Hund's rule is violated for larger $U/t>4.2$. However, due to the presence of the geometrical frustration, a systematic weakness of QMC simulation was also recognized, for both large $U/t$ range where the sign problem becomes significantly worse and small $U/t$ range where the ground state is a spin multiplet state.

\begin{figure}
\centerline{
    \includegraphics[height=2.0in,width=3.6in] {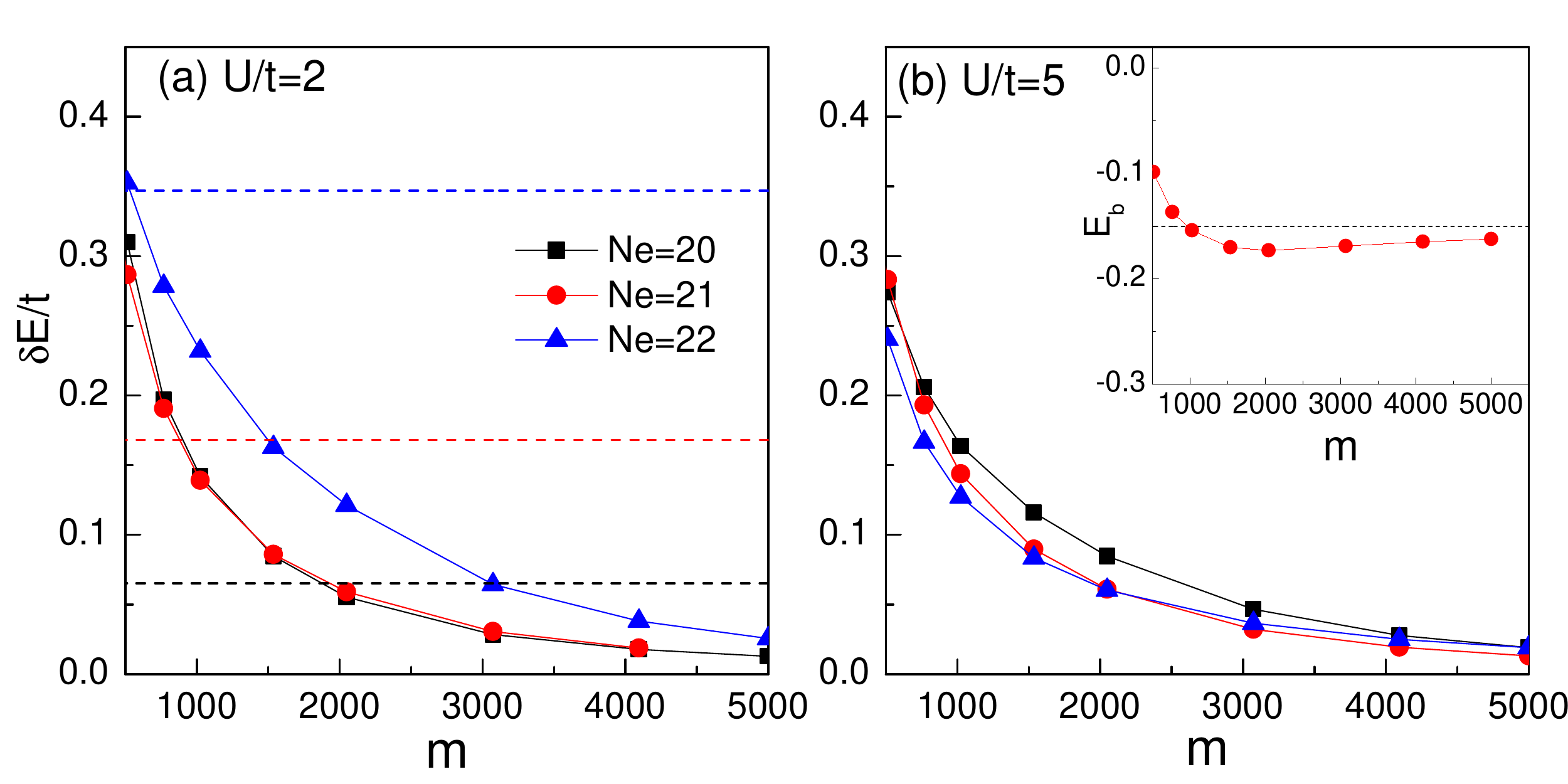}
    }
\caption{(Color online) Ground state energy difference $\delta E=E_0(m)-E_0$ as a function of DMRG states $m$ for the one-band Hubbard model (see Eq.(\ref{Eq:Hubbard})) with $N_e$ electrons on the $C_{20}$ molecule at $U/t=2$ in (a) and $U/t=5$ in (b). Here $E_0$ is the ground state energy obtained by exact diagonalization (see Ref. \cite{Berlinsky2007, Berlinsky2007p2}), while  $E_0(m)$ is the ground state energy obtained by DMRG simulation with minimal value of $z$-component total spin $S_z=s_{min}$, i.e., $s_{min}=0$ for $N_e=20$, $22$, and $s_{min}=1/2$ for $N_e=21$. The dashed lines in (a) denote the ground state energy difference between QMC and exact diagonazation with same $N_e$ and $S_z$, for comparison with the DMRG simulation labeled by the same color. The inset in (b) is the electronic pair-binding energies $E_b=2E_0(N_e=21)-E_0(N_e=22)-E_0(N_e=20)$ from DMRG (red circle) and ED simulations (solid line). Note that here we use a different definition of the pair-binding energy compared with Ref. \cite{Berlinsky2007, Berlinsky2007p2}, and a negative $E_b$ means a repulsive interaction between doped electrons.} \label{Fig:C20Hubbard}
\end{figure}

To demonstrate the reliability of the DMRG simulation, we have benchmarked the DMRG method on the one-band Hubbard model on the $C_{20}$ molecule (see Eq.(Eq:Hubbard)) by comparing the QMC and DMRG data. As we will see in Fig. \ref{Fig:C20Hubbard}, for both ranges of $U/t$, we find that the DMRG results converges very rapidly to the ED results with relatively small number of DMRG states $m$. In particular, we can get values for both the ground state energy and the pair binding energy $E_b$ that are more accurate than those obtained using QMC. \cite{Berlinsky2007, Berlinsky2007p2} For instance, with the same $N_e$ and $S_z$, DMRG can easily produce a better ground state energy than QMC with only a moderately number of DMRG states $m$, say $m\sim 1800$ for $N_e=20$ while a much smaller value $m\sim 500$ for $N_e=22$. Compared with $U/t=2$, where QMC already has a sign problem for the non-bipartite dodecahedral molecular geometry, a larger $U/t=5$ introduce significantly more sources of negative probability weight, lowering the average value of the sign, thus making that a reliable QMC simulation is not applicable\cite{Berlinsky2007, Berlinsky2007p2}. On the contrary, DMRG is immune to such a problem and still provides us with reliable results, including the ground state energy and electronic pair-binding energy $E_b(N_e=21)$ as shown in Fig.\ref{Fig:C20Hubbard}(b). Especially, a relatively small number of state $m\sim 1000$ has had given us a reliable $E_b$ which is very close to the ED results. Therefore, DMRG method works well for the Hubbard model on the $C_{20}$ molecule.

\begin{figure}
\centerline{
    \includegraphics[height=2.0in,width=3.6in] {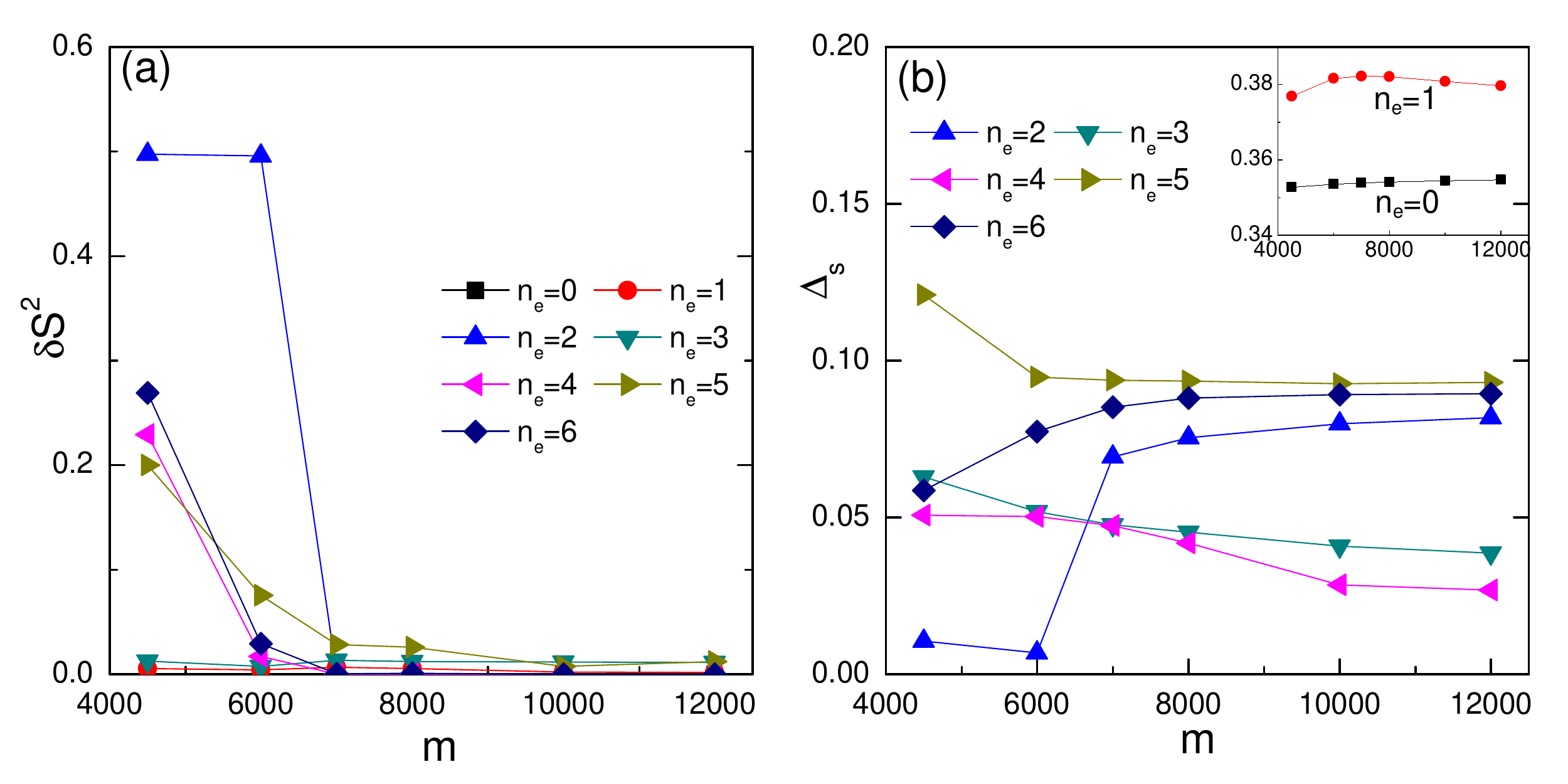}
    }
\caption{(Color online) (a)``Excess spin" $\delta S^2=\vec{S}\cdot\vec{S}-s_{min}(s_{min}+1)$ where $\vec{S}=\sum_i\vec{S_i}$ and (b) ``Spin gap" $\Delta_s=E_0(s_{min}+1)-E_0(s_{min})$ as a function of number of DMRG states $m$, for $t/J=2$ and different number of added electrons $n_e$. Here $E_0(S_z)$ is the ground state energy for given value of spin $S_z$. Inset in (b): ``Spin gap" $\Delta_s$ as a function of $m$ for $n_e=0$ and $1$.} \label{Fig:STot2SGap}
\end{figure}

\section{$t$-$J$ model on the $C_{60}$ molecule}\label{Sec:C60TJModel}

In the main text, we have introduced the $t$-$J$ model on the $C_{60}$ molecule and summarized the main DMRG results, including spin excitation gap and electronic pair-binding energy. Now, we will show more details of the DMRG simulation about the convergence of the DMRG results. For this purpose, we first consider $t/J=2$ as an example and the results are given in Fig.\ref{Fig:STot2SGap}. As seen in the figure, ``excess spin" $\delta S^2>0$ (see main text for details) when $m$ is small, indicating that the DMRG simulation may get stuck in a metastable spin multiplet state. This is because the states with smaller values of $|S_z|$ may mix with higher-lying states that have the same value of $|S_z|$ but different total spin. However, such a state is not the true ground state, instead the true ground state is obtained when $\geq 7000$, where $S_T=0$ for $n_e=0\sim 6$ cases. Therefore, the ground state is a spin singlet state, which violates the Hund's rule.

In addition to ``excess spin" $\delta S^2$, we have also calculated the 	``spin gap" $\Delta_s$ (see main text for details). Fig.\ref{Fig:STot2SGap} (b) shows the spin gap $\Delta_s$ as a function of DMRG states $m$. Similar with $\delta S^2$, $\Delta_s$ starts to converge and saturate to finite values when $m\geq 7000$, for $n_e=0\ - \ 6$. On the contrary, for the metastable state when $m\leq 6000$, $\Delta_s$ vanishes for $n_e=2$, indicating that the metastable state is a spin multiplet state. Consistent with the minimal spin state, a finite ``spin gap" $\Delta_s$ again indicates that the ground state is a minimal spin state. For $n_e=2\ - \ 4$, this indicates that Hund's rule is violated, which is in contraction to the QMC results. \cite{Berlinsky2005}

Until now, we have demonstrated that the ground state of the $t$-$J$ model on the $C_{60}$ molecule is a minimal spin state. To provide more information of the ground state, we have also computed the ``orbital gap" $\Delta_o=E_1(s_{min})-E_0(s_{min})$, where $E_0$ is the ground state energy and $E_1$ is the first excited state state, both in the same spin $S_z=s_{min}$ sector. The results are given in Fig.\ref{Fig:SGap}(b). Similar with$\Delta_s$, $\Delta_o$ is also finite for $n_e=0$, $2$ and $4$, indicating a unique ground state without orbital degeneracy. Interestingly, for other $n_e$ cases, $\Delta_o\sim 0$, indicating that the ground state is an orbital multiplet, which is consistent with \cite{Kivelson1991}.

\begin{figure}
\centerline{
    \includegraphics[height=2.4in,width=3.2in] {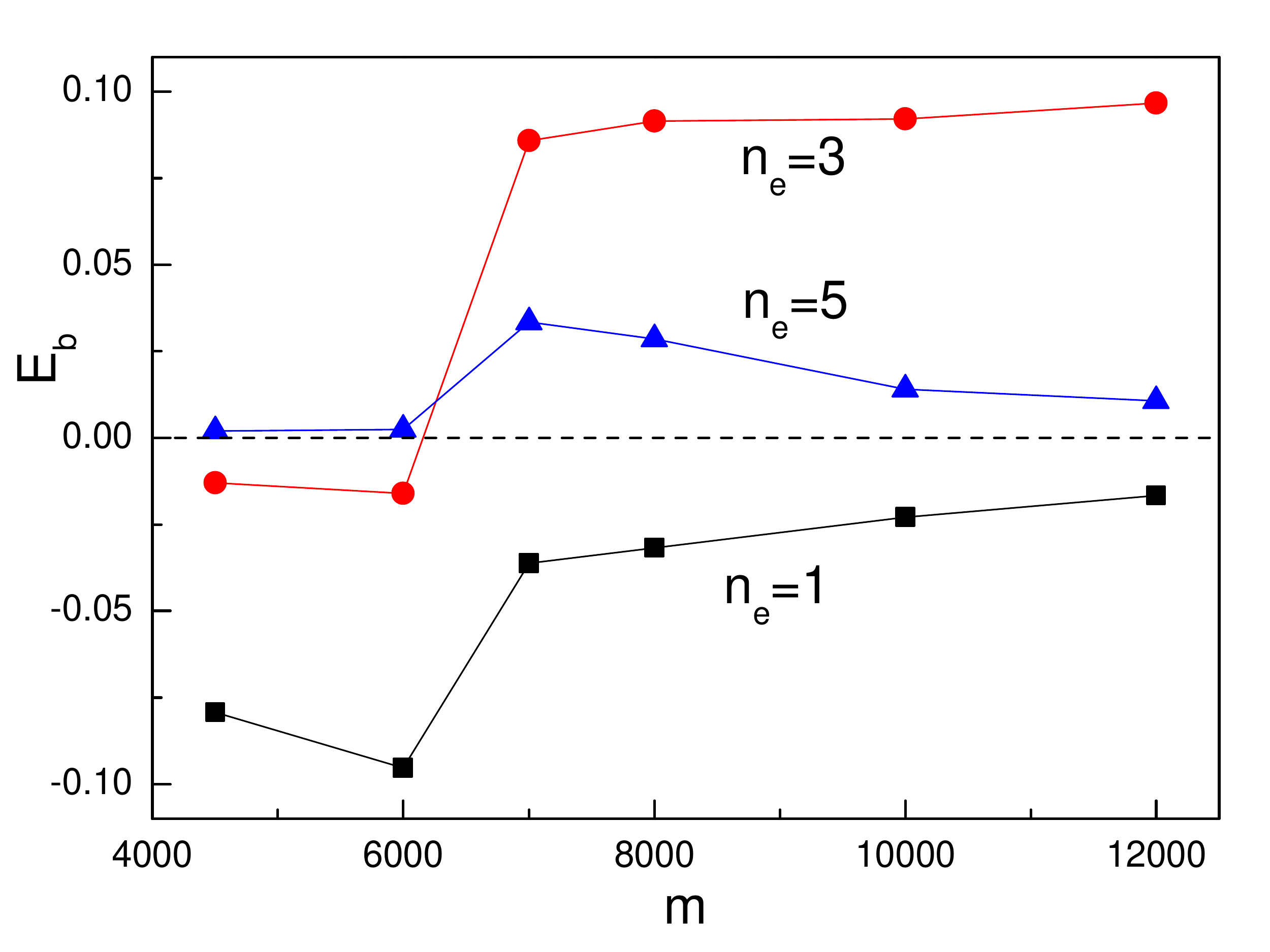}
    }
\caption{(Color online) Electronic pair-binding energy $E_b(n_e)$ for $t/J=2$ as a function of DMRG states $m$ with different number of added electrons $n_e$. The dashed line indicates the zero. The lines connecting the data points are guides to the eye only.} \label{Fig:S3Eb}
\end{figure}

Fig.\ref{Fig:S3Eb} shows the results of the electronic pair-binding energy $E_b(n_e)$ (see main text for details) for $t/J=2$ with $n_e=1$, $3$ and $5$, as a function of the number of DMRG states $m$. Similar with $\delta S^2$ and ``spin gap" $\Delta_s$, $E_b$ is also affected by the convergent problem for the numerical simulation when $m\leq 6000$, where $E_b$ is either vanishingly small or negative (repulsive interaction between doped electrons). Interestingly, when $m\geq 7000$, our DMRG simulation is well converged, and we find a big and positive $E_b(n_e=3)\sim 0.1J$. This suggests that it is energetically favorable to put four electrons on one $C_{60}$ and two on the other (positive pair binding) that putting three electrons on each of the $C_{60}$ molecules (negative pair binding). On the contrary, for both $n_e=1$ and $n_e=5$ cases, $E_b(n_e)$ is either zero or slightly negative, indicating that there are no attractive interaction between doped electrons.




\begin{thebibliography}{34}
\expandafter\ifx\csname natexlab\endcsname\relax\def\natexlab#1{#1}\fi
\expandafter\ifx\csname bibnamefont\endcsname\relax
  \def\bibnamefont#1{#1}\fi
\expandafter\ifx\csname bibfnamefont\endcsname\relax
  \def\bibfnamefont#1{#1}\fi
\expandafter\ifx\csname citenamefont\endcsname\relax
  \def\citenamefont#1{#1}\fi
\expandafter\ifx\csname url\endcsname\relax
  \def\url#1{\texttt{#1}}\fi
\expandafter\ifx\csname urlprefix\endcsname\relax\def\urlprefix{URL }\fi
\providecommand{\bibinfo}[2]{#2}
\providecommand{\eprint}[2][]{\url{#2}}

\bibitem[{\citenamefont{Hebard et~al.}(1991)\citenamefont{Hebard, Rosseinsky,
  Haddon, Murphy, Glarum, Palstra, Ramirez, and Kortan}}]{Hebard1991}
\bibinfo{author}{\bibfnamefont{A.~F.} \bibnamefont{Hebard}},
  \bibinfo{author}{\bibfnamefont{M.~J.} \bibnamefont{Rosseinsky}},
  \bibinfo{author}{\bibfnamefont{R.~C.} \bibnamefont{Haddon}},
  \bibinfo{author}{\bibfnamefont{D.~W.} \bibnamefont{Murphy}},
  \bibinfo{author}{\bibfnamefont{S.~H.} \bibnamefont{Glarum}},
  \bibinfo{author}{\bibfnamefont{T.~T.~M.} \bibnamefont{Palstra}},
  \bibinfo{author}{\bibfnamefont{A.~P.} \bibnamefont{Ramirez}},
  \bibnamefont{and} \bibinfo{author}{\bibfnamefont{A.~R.}
  \bibnamefont{Kortan}}, \bibinfo{journal}{Nature (London)}
  \textbf{\bibinfo{volume}{80}}, \bibinfo{pages}{600} (\bibinfo{year}{1991}).

\bibitem[{\citenamefont{Rosseinsky et~al.}(1991)\citenamefont{Rosseinsky,
  Ramirez, Glarum, Murphy, Haddon, Hebard, Palstra, Kortan, Zahurak, and
  Makhija}}]{Rosseinsky1991}
\bibinfo{author}{\bibfnamefont{M.~J.} \bibnamefont{Rosseinsky}},
  \bibinfo{author}{\bibfnamefont{A.~P.} \bibnamefont{Ramirez}},
  \bibinfo{author}{\bibfnamefont{S.~H.} \bibnamefont{Glarum}},
  \bibinfo{author}{\bibfnamefont{D.~W.} \bibnamefont{Murphy}},
  \bibinfo{author}{\bibfnamefont{R.~C.} \bibnamefont{Haddon}},
  \bibinfo{author}{\bibfnamefont{A.~F.} \bibnamefont{Hebard}},
  \bibinfo{author}{\bibfnamefont{T.~T.~M.} \bibnamefont{Palstra}},
  \bibinfo{author}{\bibfnamefont{A.~R.} \bibnamefont{Kortan}},
  \bibinfo{author}{\bibfnamefont{S.~M.} \bibnamefont{Zahurak}},
  \bibnamefont{and} \bibinfo{author}{\bibfnamefont{A.~V.}
  \bibnamefont{Makhija}}, \bibinfo{journal}{Phys. Rev. Lett.}
  \textbf{\bibinfo{volume}{66}}, \bibinfo{pages}{2830} (\bibinfo{year}{1991}).

\bibitem[{\citenamefont{Holczer et~al.}(1991)\citenamefont{Holczer, Klein,
  Huang, Kaner, Fu, Whetten, and Diederich}}]{Holczer1991}
\bibinfo{author}{\bibfnamefont{K.}~\bibnamefont{Holczer}},
  \bibinfo{author}{\bibfnamefont{O.}~\bibnamefont{Klein}},
  \bibinfo{author}{\bibfnamefont{S.}~\bibnamefont{Huang}},
  \bibinfo{author}{\bibfnamefont{R.~B.} \bibnamefont{Kaner}},
  \bibinfo{author}{\bibfnamefont{K.~J.} \bibnamefont{Fu}},
  \bibinfo{author}{\bibfnamefont{R.~L.} \bibnamefont{Whetten}},
  \bibnamefont{and}
  \bibinfo{author}{\bibfnamefont{F.}~\bibnamefont{Diederich}},
  \bibinfo{journal}{Science} \textbf{\bibinfo{volume}{252}},
  \bibinfo{pages}{1154} (\bibinfo{year}{1991}).

\bibitem[{\citenamefont{Gunnarsson}(1997)}]{Gunnarsson1997}
\bibinfo{author}{\bibfnamefont{O.}~\bibnamefont{Gunnarsson}},
  \bibinfo{journal}{Rev. Mod. Phys.} \textbf{\bibinfo{volume}{69}},
  \bibinfo{pages}{575} (\bibinfo{year}{1997}).

\bibitem[{\citenamefont{Ganin et~al.}(2008)\citenamefont{Ganin, Takabayashi,
  Khimyak, Margadonna, Tamai, Rosseinsky, and Prassides}}]{Prassides2008}
\bibinfo{author}{\bibfnamefont{A.~Y.} \bibnamefont{Ganin}},
  \bibinfo{author}{\bibfnamefont{Y.}~\bibnamefont{Takabayashi}},
  \bibinfo{author}{\bibfnamefont{Y.~Z.} \bibnamefont{Khimyak}},
  \bibinfo{author}{\bibfnamefont{S.}~\bibnamefont{Margadonna}},
  \bibinfo{author}{\bibfnamefont{A.}~\bibnamefont{Tamai}},
  \bibinfo{author}{\bibfnamefont{M.~J.} \bibnamefont{Rosseinsky}},
  \bibnamefont{and}
  \bibinfo{author}{\bibfnamefont{K.}~\bibnamefont{Prassides}},
  \bibinfo{journal}{Nat. Mat.} \textbf{\bibinfo{volume}{7}},
  \bibinfo{pages}{367} (\bibinfo{year}{2008}).

\bibitem[{\citenamefont{Takabayashi et~al.}(2009)\citenamefont{Takabayashi,
  Ganin, Jeglic, Arcon, Takano, Iwasa, Ohishi, Takata, Takeshita, Prassides
  et~al.}}]{Takabayashi2009}
\bibinfo{author}{\bibfnamefont{Y.}~\bibnamefont{Takabayashi}},
  \bibinfo{author}{\bibfnamefont{A.~Y.} \bibnamefont{Ganin}},
  \bibinfo{author}{\bibfnamefont{P.}~\bibnamefont{Jeglic}},
  \bibinfo{author}{\bibfnamefont{D.}~\bibnamefont{Arcon}},
  \bibinfo{author}{\bibfnamefont{T.}~\bibnamefont{Takano}},
  \bibinfo{author}{\bibfnamefont{Y.}~\bibnamefont{Iwasa}},
  \bibinfo{author}{\bibfnamefont{Y.}~\bibnamefont{Ohishi}},
  \bibinfo{author}{\bibfnamefont{M.}~\bibnamefont{Takata}},
  \bibinfo{author}{\bibfnamefont{N.}~\bibnamefont{Takeshita}},
  \bibinfo{author}{\bibfnamefont{K.}~\bibnamefont{Prassides}},
  \bibnamefont{et~al.}, \bibinfo{journal}{Science}
  \textbf{\bibinfo{volume}{323}}, \bibinfo{pages}{1585} (\bibinfo{year}{2009}).

\bibitem[{\citenamefont{Ganin et~al.}(2010)\citenamefont{Ganin, Takabayashi,
  Jeglic, Arcon, Potocnik, Baker, Ohishi, McDonald, Tzirakis, McLennan
  et~al.}}]{Prassides2010}
\bibinfo{author}{\bibfnamefont{A.~Y.} \bibnamefont{Ganin}},
  \bibinfo{author}{\bibfnamefont{Y.}~\bibnamefont{Takabayashi}},
  \bibinfo{author}{\bibfnamefont{P.}~\bibnamefont{Jeglic}},
  \bibinfo{author}{\bibfnamefont{D.}~\bibnamefont{Arcon}},
  \bibinfo{author}{\bibfnamefont{A.}~\bibnamefont{Potocnik}},
  \bibinfo{author}{\bibfnamefont{P.~J.} \bibnamefont{Baker}},
  \bibinfo{author}{\bibfnamefont{Y.}~\bibnamefont{Ohishi}},
  \bibinfo{author}{\bibfnamefont{M.~T.} \bibnamefont{McDonald}},
  \bibinfo{author}{\bibfnamefont{M.~D.} \bibnamefont{Tzirakis}},
  \bibinfo{author}{\bibfnamefont{A.}~\bibnamefont{McLennan}},
  \bibnamefont{et~al.}, \bibinfo{journal}{Nature}
  \textbf{\bibinfo{volume}{466}}, \bibinfo{pages}{221} (\bibinfo{year}{2010}).

\bibitem[{\citenamefont{Ihara et~al.}(2010)\citenamefont{Ihara, Alloul,
  Wzietek, Pontiroli, Mazzani, and Ricco}}]{Alloul2010}
\bibinfo{author}{\bibfnamefont{Y.}~\bibnamefont{Ihara}},
  \bibinfo{author}{\bibfnamefont{H.}~\bibnamefont{Alloul}},
  \bibinfo{author}{\bibfnamefont{P.}~\bibnamefont{Wzietek}},
  \bibinfo{author}{\bibfnamefont{D.}~\bibnamefont{Pontiroli}},
  \bibinfo{author}{\bibfnamefont{M.}~\bibnamefont{Mazzani}}, \bibnamefont{and}
  \bibinfo{author}{\bibfnamefont{M.}~\bibnamefont{Ricco}},
  \bibinfo{journal}{Phys. Rev. Lett.} \textbf{\bibinfo{volume}{104}}
  (\bibinfo{year}{2010}).

\bibitem[{\citenamefont{Wzietek et~al.}(2014)\citenamefont{Wzietek, Mito,
  Alloul, Pontiroli, Aramini, and Ricco}}]{Alloul2014}
\bibinfo{author}{\bibfnamefont{P.}~\bibnamefont{Wzietek}},
  \bibinfo{author}{\bibfnamefont{T.}~\bibnamefont{Mito}},
  \bibinfo{author}{\bibfnamefont{H.}~\bibnamefont{Alloul}},
  \bibinfo{author}{\bibfnamefont{D.}~\bibnamefont{Pontiroli}},
  \bibinfo{author}{\bibfnamefont{M.}~\bibnamefont{Aramini}}, \bibnamefont{and}
  \bibinfo{author}{\bibfnamefont{M.}~\bibnamefont{Ricco}},
  \bibinfo{journal}{Phys. Rev. Lett.} \textbf{\bibinfo{volume}{112}}
  (\bibinfo{year}{2014}).

\bibitem[{\citenamefont{Chakravarty
  et~al.}(1991{\natexlab{a}})\citenamefont{Chakravarty, Gelfand, and
  Kivelson}}]{Kivelson1991}
\bibinfo{author}{\bibfnamefont{S.}~\bibnamefont{Chakravarty}},
  \bibinfo{author}{\bibfnamefont{M.~P.} \bibnamefont{Gelfand}},
  \bibnamefont{and} \bibinfo{author}{\bibfnamefont{S.}~\bibnamefont{Kivelson}},
  \bibinfo{journal}{Science} \textbf{\bibinfo{volume}{254}},
  \bibinfo{pages}{970} (\bibinfo{year}{1991}{\natexlab{a}}).

\bibitem[{\citenamefont{Chakravarty and Kivelson}(1991)}]{Kivelson1991p2}
\bibinfo{author}{\bibfnamefont{S.}~\bibnamefont{Chakravarty}} \bibnamefont{and}
  \bibinfo{author}{\bibfnamefont{S.}~\bibnamefont{Kivelson}},
  \bibinfo{journal}{Europhys. Lett.} \textbf{\bibinfo{volume}{16}},
  \bibinfo{pages}{751} (\bibinfo{year}{1991}).

\bibitem[{\citenamefont{White et~al.}(1992)\citenamefont{White, Chakravarty,
  Gelfand, and Kivelson}}]{Kivelson1992}
\bibinfo{author}{\bibfnamefont{S.~R.} \bibnamefont{White}},
  \bibinfo{author}{\bibfnamefont{S.}~\bibnamefont{Chakravarty}},
  \bibinfo{author}{\bibfnamefont{M.~P.} \bibnamefont{Gelfand}},
  \bibnamefont{and} \bibinfo{author}{\bibfnamefont{S.~A.}
  \bibnamefont{Kivelson}}, \bibinfo{journal}{Phys. Rev. B}
  \textbf{\bibinfo{volume}{45}}, \bibinfo{pages}{5062} (\bibinfo{year}{1992}).

\bibitem[{\citenamefont{Murthy and Auerbach}(1992)}]{Murthy1992}
\bibinfo{author}{\bibfnamefont{G.~N.} \bibnamefont{Murthy}} \bibnamefont{and}
  \bibinfo{author}{\bibfnamefont{A.}~\bibnamefont{Auerbach}},
  \bibinfo{journal}{Phys. Rev. B} \textbf{\bibinfo{volume}{46}},
  \bibinfo{pages}{331} (\bibinfo{year}{1992}).

\bibitem[{\citenamefont{Goff and Phillips}(1992)}]{Goff1992}
\bibinfo{author}{\bibfnamefont{W.~E.} \bibnamefont{Goff}} \bibnamefont{and}
  \bibinfo{author}{\bibfnamefont{P.}~\bibnamefont{Phillips}},
  \bibinfo{journal}{Phys. Rev. B} \textbf{\bibinfo{volume}{46}},
  \bibinfo{pages}{603} (\bibinfo{year}{1992}).

\bibitem[{\citenamefont{Baskaran}(1992)}]{Baskaran1992}
\bibinfo{author}{\bibfnamefont{G.}~\bibnamefont{Baskaran}},
  \bibinfo{journal}{Indian Journal of Chemistry} \textbf{\bibinfo{volume}{31}},
  \bibinfo{pages}{F104} (\bibinfo{year}{1992}).

\bibitem[{\citenamefont{Scalettar et~al.}(1993)\citenamefont{Scalettar, Moreo,
  Dagotto, Bergomi, Jolicoeur, and Monien}}]{Scalettar1993}
\bibinfo{author}{\bibfnamefont{R.~T.} \bibnamefont{Scalettar}},
  \bibinfo{author}{\bibfnamefont{A.}~\bibnamefont{Moreo}},
  \bibinfo{author}{\bibfnamefont{E.}~\bibnamefont{Dagotto}},
  \bibinfo{author}{\bibfnamefont{L.}~\bibnamefont{Bergomi}},
  \bibinfo{author}{\bibfnamefont{T.}~\bibnamefont{Jolicoeur}},
  \bibnamefont{and} \bibinfo{author}{\bibfnamefont{H.}~\bibnamefont{Monien}},
  \bibinfo{journal}{Phys. Rev. B} \textbf{\bibinfo{volume}{47}},
  \bibinfo{pages}{12316} (\bibinfo{year}{1993}).

\bibitem[{\citenamefont{Goff and Phillips}(1993)}]{Goff1993}
\bibinfo{author}{\bibfnamefont{W.~E.} \bibnamefont{Goff}} \bibnamefont{and}
  \bibinfo{author}{\bibfnamefont{P.}~\bibnamefont{Phillips}},
  \bibinfo{journal}{Phys. Rev. B} \textbf{\bibinfo{volume}{48}},
  \bibinfo{pages}{3491} (\bibinfo{year}{1993}).

\bibitem[{\citenamefont{Krivnov et~al.}(1994)\citenamefont{Krivnov, Shamovsky,
  Tornau, and Rosengren}}]{Krivnov1994}
\bibinfo{author}{\bibfnamefont{V.~Y.} \bibnamefont{Krivnov}},
  \bibinfo{author}{\bibfnamefont{I.~L.} \bibnamefont{Shamovsky}},
  \bibinfo{author}{\bibfnamefont{E.~E.} \bibnamefont{Tornau}},
  \bibnamefont{and}
  \bibinfo{author}{\bibfnamefont{A.}~\bibnamefont{Rosengren}},
  \bibinfo{journal}{Phys. Rev. B} \textbf{\bibinfo{volume}{50}},
  \bibinfo{pages}{12144} (\bibinfo{year}{1994}).

\bibitem[{\citenamefont{Zhang et~al.}(1995)\citenamefont{Zhang, Ma, Sun, Lee,
  and Park}}]{Zhang1995}
\bibinfo{author}{\bibfnamefont{G.~P.} \bibnamefont{Zhang}},
  \bibinfo{author}{\bibfnamefont{Y.~S.} \bibnamefont{Ma}},
  \bibinfo{author}{\bibfnamefont{X.}~\bibnamefont{Sun}},
  \bibinfo{author}{\bibfnamefont{K.~H.} \bibnamefont{Lee}}, \bibnamefont{and}
  \bibinfo{author}{\bibfnamefont{T.~Y.} \bibnamefont{Park}},
  \bibinfo{journal}{Phys. Rev. B} \textbf{\bibinfo{volume}{52}},
  \bibinfo{pages}{6081} (\bibinfo{year}{1995}).

\bibitem[{\citenamefont{Sondhi et~al.}(1995)\citenamefont{Sondhi, Gelfand, Lin,
  and Campbell}}]{Sondhi1995}
\bibinfo{author}{\bibfnamefont{S.~L.} \bibnamefont{Sondhi}},
  \bibinfo{author}{\bibfnamefont{M.~P.} \bibnamefont{Gelfand}},
  \bibinfo{author}{\bibfnamefont{H.~Q.} \bibnamefont{Lin}}, \bibnamefont{and}
  \bibinfo{author}{\bibfnamefont{D.~K.} \bibnamefont{Campbell}},
  \bibinfo{journal}{Phys. Rev. B} \textbf{\bibinfo{volume}{51}},
  \bibinfo{pages}{5943} (\bibinfo{year}{1995}).

\bibitem[{\citenamefont{Campbell et~al.}(1995)\citenamefont{Campbell, Gelfand,
  Lin, and Sondhi}}]{Campbell1995}
\bibinfo{author}{\bibfnamefont{D.~K.} \bibnamefont{Campbell}},
  \bibinfo{author}{\bibfnamefont{M.~P.} \bibnamefont{Gelfand}},
  \bibinfo{author}{\bibfnamefont{H.~Q.} \bibnamefont{Lin}}, \bibnamefont{and}
  \bibinfo{author}{\bibfnamefont{S.~L.} \bibnamefont{Sondhi}},
  \bibinfo{journal}{Synthetic Metals} \textbf{\bibinfo{volume}{70}},
  \bibinfo{pages}{1523} (\bibinfo{year}{1995}).

\bibitem[{\citenamefont{Lin et~al.}(2005{\natexlab{a}})\citenamefont{Lin,
  \ifmmode~\check{S}\else \v{S}\fi{}makov, S\o{}rensen, Kallin, and
  Berlinsky}}]{Kallin2005}
\bibinfo{author}{\bibfnamefont{F.}~\bibnamefont{Lin}},
  \bibinfo{author}{\bibfnamefont{J.}~\bibnamefont{\ifmmode~\check{S}\else
  \v{S}\fi{}makov}}, \bibinfo{author}{\bibfnamefont{E.~S.}
  \bibnamefont{S\o{}rensen}},
  \bibinfo{author}{\bibfnamefont{C.}~\bibnamefont{Kallin}}, \bibnamefont{and}
  \bibinfo{author}{\bibfnamefont{A.~J.} \bibnamefont{Berlinsky}},
  \bibinfo{journal}{Phys. Rev. B} \textbf{\bibinfo{volume}{71}},
  \bibinfo{pages}{165436} (\bibinfo{year}{2005}{\natexlab{a}}).

\bibitem[{\citenamefont{White}(1992)}]{White1992DMRG}
\bibinfo{author}{\bibfnamefont{S.~R.} \bibnamefont{White}},
  \bibinfo{journal}{Phys. Rev. Lett.} \textbf{\bibinfo{volume}{69}},
  \bibinfo{pages}{2863} (\bibinfo{year}{1992}).

\bibitem[{\citenamefont{Schollw\"ock}(2005)}]{DMRG_RMP}
\bibinfo{author}{\bibfnamefont{U.}~\bibnamefont{Schollw\"ock}},
  \bibinfo{journal}{Rev. Mod. Phys.} \textbf{\bibinfo{volume}{77}},
  \bibinfo{pages}{259} (\bibinfo{year}{2005}).

\bibitem[{\citenamefont{Stoudenmire and White}(2002)}]{DMRG_2D}
\bibinfo{author}{\bibfnamefont{E.~M.} \bibnamefont{Stoudenmire}}
  \bibnamefont{and} \bibinfo{author}{\bibfnamefont{S.~R.} \bibnamefont{White}},
  \bibinfo{journal}{Annual Review of Condensed Matter Physics}
  \textbf{\bibinfo{volume}{3}}, \bibinfo{pages}{111} (\bibinfo{year}{2002}).

\bibitem[{\citenamefont{Haddon et~al.}(1986)\citenamefont{Haddon, Brus, and
  Raghavachari}}]{Haddon1986}
\bibinfo{author}{\bibfnamefont{R.~C.} \bibnamefont{Haddon}},
  \bibinfo{author}{\bibfnamefont{L.~E.} \bibnamefont{Brus}}, \bibnamefont{and}
  \bibinfo{author}{\bibfnamefont{K.}~\bibnamefont{Raghavachari}},
  \bibinfo{journal}{Chem. Phys. Lett.} \textbf{\bibinfo{volume}{125}},
  \bibinfo{pages}{459} (\bibinfo{year}{1986}).

\bibitem[{\citenamefont{Coffey and Trugman}(1992)}]{Trugman1992}
\bibinfo{author}{\bibfnamefont{D.}~\bibnamefont{Coffey}} \bibnamefont{and}
  \bibinfo{author}{\bibfnamefont{S.~A.} \bibnamefont{Trugman}},
  \bibinfo{journal}{Phys. Rev. Lett.} \textbf{\bibinfo{volume}{69}},
  \bibinfo{pages}{176} (\bibinfo{year}{1992}).

\bibitem[{\citenamefont{Chakravarty
  et~al.}(1991{\natexlab{b}})\citenamefont{Chakravarty, Chayes, and
  Kivelson}}]{chayes}
\bibinfo{author}{\bibfnamefont{S.}~\bibnamefont{Chakravarty}},
  \bibinfo{author}{\bibfnamefont{L.}~\bibnamefont{Chayes}}, \bibnamefont{and}
  \bibinfo{author}{\bibfnamefont{S.~A.} \bibnamefont{Kivelson}},
  \bibinfo{journal}{Letters in Mathematical Physics}
  \textbf{\bibinfo{volume}{23}}, \bibinfo{pages}{265}
  (\bibinfo{year}{1991}{\natexlab{b}}).

\bibitem[{\citenamefont{Auerbach et~al.}(1994)\citenamefont{Auerbach, Manini,
  and Tosatti}}]{auerbach}
\bibinfo{author}{\bibfnamefont{A.}~\bibnamefont{Auerbach}},
  \bibinfo{author}{\bibfnamefont{N.}~\bibnamefont{Manini}}, \bibnamefont{and}
  \bibinfo{author}{\bibfnamefont{E.}~\bibnamefont{Tosatti}},
  \bibinfo{journal}{Phys. Rev. B} \textbf{\bibinfo{volume}{49}},
  \bibinfo{pages}{12998} (\bibinfo{year}{1994}).

\bibitem[{\citenamefont{Capone et~al.}(2009)\citenamefont{Capone, Fabrizio,
  Castellani, and Tosatti}}]{tosatti}
\bibinfo{author}{\bibfnamefont{M.}~\bibnamefont{Capone}},
  \bibinfo{author}{\bibfnamefont{M.}~\bibnamefont{Fabrizio}},
  \bibinfo{author}{\bibfnamefont{C.}~\bibnamefont{Castellani}},
  \bibnamefont{and} \bibinfo{author}{\bibfnamefont{E.}~\bibnamefont{Tosatti}},
  \bibinfo{journal}{Rev. Mod. Phys.} \textbf{\bibinfo{volume}{81}},
  \bibinfo{pages}{943} (\bibinfo{year}{2009}).

\bibitem[{\citenamefont{Zadik et~al.}(2015)\citenamefont{Zadik, Takabayashi,
  andR. H.~Colman, Ganin, Potocnik, Jeglic, D.Arcon, Matus, Kamaras, Kasahara
  et~al.}}]{prassides}
\bibinfo{author}{\bibfnamefont{R.~H.} \bibnamefont{Zadik}},
  \bibinfo{author}{\bibfnamefont{Y.}~\bibnamefont{Takabayashi}},
  \bibinfo{author}{\bibfnamefont{G.~K.} \bibnamefont{andR. H.~Colman}},
  \bibinfo{author}{\bibfnamefont{A.~Y.} \bibnamefont{Ganin}},
  \bibinfo{author}{\bibfnamefont{A.}~\bibnamefont{Potocnik}},
  \bibinfo{author}{\bibfnamefont{P.}~\bibnamefont{Jeglic}},
  \bibinfo{author}{\bibnamefont{D.Arcon}},
  \bibinfo{author}{\bibfnamefont{P.}~\bibnamefont{Matus}},
  \bibinfo{author}{\bibfnamefont{K.}~\bibnamefont{Kamaras}},
  \bibinfo{author}{\bibfnamefont{Y.}~\bibnamefont{Kasahara}},
  \bibnamefont{et~al.}, \bibinfo{journal}{Sci. Adv.}
  \textbf{\bibinfo{volume}{1}}, \bibinfo{pages}{e1500059}
  (\bibinfo{year}{2015}).

\bibitem[{\citenamefont{Lin et~al.}(2007{\natexlab{a}})\citenamefont{Lin,
  S\o{}rensen, Kallin, and Berlinsky}}]{Berlinsky2007}
\bibinfo{author}{\bibfnamefont{F.}~\bibnamefont{Lin}},
  \bibinfo{author}{\bibfnamefont{E.~S.} \bibnamefont{S\o{}rensen}},
  \bibinfo{author}{\bibfnamefont{C.}~\bibnamefont{Kallin}}, \bibnamefont{and}
  \bibinfo{author}{\bibfnamefont{A.~J.} \bibnamefont{Berlinsky}},
  \bibinfo{journal}{Phys. Rev. B} \textbf{\bibinfo{volume}{76}},
  \bibinfo{pages}{033414} (\bibinfo{year}{2007}{\natexlab{a}}).

\bibitem[{\citenamefont{Lin et~al.}(2007{\natexlab{b}})\citenamefont{Lin,
  S\o{}rensen, Kallin, and Berlinsky}}]{Berlinsky2007p2}
\bibinfo{author}{\bibfnamefont{F.}~\bibnamefont{Lin}},
  \bibinfo{author}{\bibfnamefont{E.~S.} \bibnamefont{S\o{}rensen}},
  \bibinfo{author}{\bibfnamefont{C.}~\bibnamefont{Kallin}}, \bibnamefont{and}
  \bibinfo{author}{\bibfnamefont{A.~J.} \bibnamefont{Berlinsky}},
  \bibinfo{journal}{J. Phys.: Condens. Matter} \textbf{\bibinfo{volume}{19}},
  \bibinfo{pages}{456206} (\bibinfo{year}{2007}{\natexlab{b}}).

\bibitem[{\citenamefont{Lin et~al.}(2005{\natexlab{b}})\citenamefont{Lin,
  \ifmmode~\check{S}\else \v{S}\fi{}makov, S\o{}rensen, Kallin, and
  Berlinsky}}]{Berlinsky2005}
\bibinfo{author}{\bibfnamefont{F.}~\bibnamefont{Lin}},
  \bibinfo{author}{\bibfnamefont{J.}~\bibnamefont{\ifmmode~\check{S}\else
  \v{S}\fi{}makov}}, \bibinfo{author}{\bibfnamefont{E.~S.}
  \bibnamefont{S\o{}rensen}},
  \bibinfo{author}{\bibfnamefont{C.}~\bibnamefont{Kallin}}, \bibnamefont{and}
  \bibinfo{author}{\bibfnamefont{A.~J.} \bibnamefont{Berlinsky}},
  \bibinfo{journal}{Phys. Rev. B} \textbf{\bibinfo{volume}{71}},
  \bibinfo{pages}{165436} (\bibinfo{year}{2005}{\natexlab{b}}).

\end{thebibliography}

\end{document}